**Tensor Properties of the Magneto-electric Coupling in Hexaferrites**

Carmine Vittoria [a], Somu Sivasubramanian [b] and Allan Widom [c], Northeastern University, Boston, Massachusetts 02115

a-Electrical and Computer Engineering Department; b-Center for High-rate Nanomanufacturing; c-Physics Department

Recent data reported on the magneto-electric coupling, $\alpha$, of hexaferrite materials is very high, and they have the potential to impact future technologies in a major way. The fundamental question addressed in this paper is what does $\alpha$ depend on and, therefore, to what extent can $\alpha$ increase in future experiments. Mathematical formulation of a microscopic theory would be rather complex, especially for a complex crystal structure as the Z-type hexaferrite, for example. We have opted for a thermodynamic argument in which a functional relationship between $\alpha$ and material parameters was derived. We find that $\alpha$ is best described as a tensor property proportional to the product of the magnetostriction and piezoelectric strain tensors. Quantitative estimates of $\alpha$ are given for hexaferrites and composites.

**Introduction**

Recently, magneto-electric (ME) effects have been reported on M [1-2], Y [3], and Z-type [4-5] hexaferrites at room temperature. The linear ME coupling parameter, $\alpha$, was measured to be $7.6 \times 10^{-10}$ s/m for polycrystalline [4] and $2.3 \times 10^{-6}$ for single crystal Z-type hexaferrites [5]. The reader is referred to ref [6] for other forms of ME coupling. Early measurements have reported values of $\alpha \approx 4 \times 10^{-12}$ s/m on $Cr_2O_3$ [7]. There has been steady progress ever since the first measurement [7]. The increase in $\alpha$ as measured recently in hexaferrites is significant for the materials can potentially impact every facet of modern technologies whereby power consumption, miniaturization, compatibility with other technologies and computing circuitry is critical or crucial. The fundamental question addressed in this paper is to what extent can $\alpha$ increase further in the future. Past theories as related to ferrites tended to be qualitative limiting themselves mostly on the physical conditions or requirements by which ferrites give rise to ME effects at any temperature.

The argument presented in various models [8-10] is that non co-linearity of spins as formulated by the Dzyaloshinskii-Morya [11], DM, interaction equation is essential in exciting ME effects in hexaferrites. According to ref [4] the substitution of Sr ion for Ba ion in Z-type hexaferrites (this applies to other hexaferrites as well) strains the bonding of the chemical combination of Fe-O-Fe located near Sr ion or the 12k site. The change of the bonding angle has two ramifications: (1) it affects the ground state of the spins globally, giving rise to spin spiral configurations. Thus, changes in strains translate into change in magnetic moment via the magnetostriction interaction. (2) The application of an electric field, E, strains the material and induces electric polarization giving rise to piezoelectric effects as well as changes in magnetic moment, the ME effect.

In summary, past studies have argued for the existence of non co-linearity of spins to explain ME effects [8-10] as manifested through the ME coupling parameter ,α. Typically, ME hexaferrites are ferrimagnetic and piezoelectric at room temperature and, as such, they may be categorized as multiferroics. To our knowledge no studies have correlated α to the piezoelectric strain coefficient, d, and/or the magnetostriction constant, λ, or any other physical parameter. This paper addresses the correlation or functional relationship between α and any other physical parameter, including d and λ. We show that indeed α is a tensor and in the simplified form it is a function of the product of d and λ.

A microscopic theory encompassing the above two points must necessarily be a global theory which includes all of the magnetic ions (including Co ions, for example) in the unit cell of a hexaferrite and not simply sites near the 12k site in order to obtain a quantitative estimate of $\alpha$. This is a very difficult undertaking which depends on the details of site occupancies for all the ions in the unit cell. In order to avoid these type of difficulties we have opted for a thermodynamic argument which does not depend on the details of microscopic calculations.

## THERMODYNAMICS

The electromagnetic Helmholtz free energy per unit volume is here defined as the function $F(\mathbf{E}, \mathbf{H}, \mathbf{u}, T)$ where $\mathbf{E}$ is the vector electric field, $\mathbf{H}$ is the vector magnetic intensity, $\mathbf{u}$ is the tensor strain and T is the temperature.

$$dF = -SdT - \mathbf{M} \cdot d\mathbf{H} - \mathbf{P} \cdot d\mathbf{E} + \sigma : d\mathbf{u}, \tag{1}$$

where S is the entropy per unit volume, $\mathbf{M}$ is the vector magnetization, $\mathbf{P}$ is the vector polarization and $\sigma$ is the tensor stress. The electromagnetic Gibbs free energy per unit volume is here defined as the function $G(\mathbf{E}, \mathbf{H}, \sigma, T)$ implicit in the Free energy minimum principle

$$G(\mathbf{E}, \mathbf{H}, \sigma, T) = \min_{\mathbf{u}} [F(\mathbf{E}, \mathbf{H}, \mathbf{u}, T) - \sigma : \mathbf{u}], \tag{2}$$

so that

$$dG = -SdT - \mathbf{M} \cdot d\mathbf{H} - \mathbf{P} \cdot d\mathbf{E} + \mathbf{u} : d\sigma \tag{3}$$

We employ Gaussian CGS units wherein

$$\mathbf{B} = \mathbf{H} + 4\pi\mathbf{M} \text{ and } \mathbf{D} = \mathbf{E} + 4\pi\mathbf{P} . \tag{4}$$

There are two magneto-electric coefficients. The first holds the strain constant and the second holds the stress constant, respectively,

$$\alpha_{ij} = \left(\frac{\partial P_i}{\partial H_j}\right)_{\mathbf{E},\mathbf{u},T} = \left(\frac{\partial M_i}{\partial E_j}\right)_{\mathbf{H},\mathbf{u},T} = -\left[\frac{\partial^2 F}{\partial E_i \partial H_j}\right]_{\mathbf{u},T} \tag{5}$$

and

$$\tilde{\alpha}_{ij} = \left(\frac{\partial P_i}{\partial H_j}\right)_{\mathbf{E},\sigma,T} = \left(\frac{\partial M_i}{\partial E_j}\right)_{\mathbf{H},\sigma,T} = -\left[\frac{\partial^2 G}{\partial E_i \partial H_j}\right]_{\sigma,T} \tag{6}$$

Of interest here is the difference

$$\Delta\alpha_{ij} = \tilde{\alpha}_{ij} - \alpha_{ij} \tag{7}$$

The total magneto-electric coefficient in eq. (6) is the sum of the intrinsic magneto-electric coefficient in eq.(5) and $\Delta\alpha_{ij}$ as in eq.(7).

The intrinsic magneto-electric coefficient may be evaluated using the differentiation chain rules

$$\alpha_{ij} = \left(\frac{\partial P_i}{\partial H_j}\right)_{\mathbf{E},\mathbf{u},T} = \left(\frac{\partial P_i}{\partial H_j}\right)_{\mathbf{E},\sigma,T} + \left(\frac{\partial P_i}{\partial \sigma_{kl}}\right)_{\mathbf{E},\mathbf{H},T} \left(\frac{\partial \sigma_{kl}}{\partial H_j}\right)_{\mathbf{E},\mathbf{u},T} \tag{8}$$

$$\alpha_{ij} = \left(\frac{\partial P_i}{\partial H_j}\right)_{\mathbf{E},\mathbf{u},T} = \tilde{\alpha}_{ij} - \left(\frac{\partial P_i}{\partial \sigma_{kl}}\right)_{\mathbf{E},\mathbf{H},T} \left(\frac{\partial \sigma_{kl}}{\partial u_{mn}}\right)_{\mathbf{E},\mathbf{H},T} \left(\frac{\partial u_{nm}}{\partial H_j}\right)_{\mathbf{E},\sigma,T} \tag{9}$$

Employing the Maxwell relation from eq.(3)

$$\left(\frac{\partial u_{nm}}{\partial H_j}\right)_{\mathbf{E},\sigma,T} = \left(\frac{\partial M_j}{\partial \sigma_{nm}}\right)_{\mathbf{E},\mathbf{H},T} . \tag{10}$$

In virtue of eq.(9) yields

$$\tilde{\alpha}_{ij} = \alpha_{ij} + \left(\frac{\partial P_i}{\partial \sigma_{kl}}\right)_{\mathbf{E},\mathbf{H},T} \left(\frac{\partial \sigma_{kl}}{\partial u_{mn}}\right)_{\mathbf{E},\mathbf{H},T} \left(\frac{\partial M_j}{\partial \sigma_{nm}}\right)_{\mathbf{E},\mathbf{H},T} . \tag{11}$$

The piezomagnetic and piezoelectric coefficients are defined by

$$\Lambda^{\mathbf{M}}_{ijk} = \left(\frac{\partial M_j}{\partial \sigma_{jk}}\right)_{\mathbf{E},\mathbf{H},T} \text{ and } \Lambda^{\mathbf{P}}_{ijk} = \left(\frac{\partial P_j}{\partial \sigma_{jk}}\right)_{\mathbf{E},\mathbf{H},T} \tag{12}$$

and the elastic modulus tensor is defined as

$$\mathcal{E}_{ijkl} = \left(\frac{\partial \sigma_{ij}}{\partial u_{kl}}\right)_{\mathbf{E},\mathbf{H},T} . \tag{13}$$

Eqs.(7), (12), and (13) yield the following

$$\tilde{\alpha}_{ij} = \alpha_{ij} + \Lambda^{\mathbf{P}}_{ikl}\mathcal{E}_{klnm}\Lambda^{\mathbf{M}}_{jnm} \tag{14}$$

$$\Delta\alpha_{ij} = \Lambda^{\mathbf{P}}_{ikl}\mathcal{E}_{klnm}\Lambda^{\mathbf{M}}_{jnm}. \tag{15}$$

The piezomagnetic coefficients for a single ferromagnetic domain as strictly defined above can be written in terms of the magnetostriction coefficients

$$a_{ijkl} = \frac{1}{2}\left(\frac{\partial^2 u_{ij}}{\partial M_k \partial M_l}\right)_{\mathbf{E},\sigma,T} \tag{16}$$

Note the limits

$$\lim_{\mathbf{H}\to \mathbf{0}}(a_{ijkl}M_l) = \frac{1}{2}\lim_{\mathbf{H}\to \mathbf{0}}\left(\frac{\partial u_{ij}}{\partial M_k}\right)_{\mathbf{E},\sigma,T} = \frac{1}{2}\lim_{\mathbf{H}\to \mathbf{0}}\left(\frac{\partial u_{ij}}{\partial H_l}\right)_{\mathbf{E},\sigma,T}\left(\frac{\partial H_l}{\partial M_k}\right)_{\mathbf{E},\sigma,T}. \quad (17)$$

If we define the magnetic susceptibility as

$$\chi_{ij}^{\mathrm{M}} = \left(\frac{\partial M_i}{\partial H_j}\right)_{\mathbf{E},\sigma,T}, \quad (18)$$

eq.(17) implies a relationship between the magnetostriction coefficients to the piezomagnetic coefficients for a single domain ferromagnet.

$$\lim_{\mathbf{H}\to \mathbf{0}}(\Lambda_{ijk}^{\mathrm{M}}) = 2\lim_{\mathbf{H}\to \mathbf{0}}(a_{ijnl}\chi_{nk}^{\mathrm{M}}M_l) \quad (19)$$

The central result is a thermodynamic identity

$$\tilde{\alpha}_{ij} = \alpha_{ij} + \Lambda_{ikl}^{\mathrm{P}}\mathcal{E}_{klnm}\Lambda_{jnm}^{\mathrm{M}} = \alpha_{ij} + \Delta\alpha_{ij}$$

A few comments are in order

(i) If the crystal were absolutely rigid so that the atoms did not move at all when fields are applied (zero strain), then the intrinsic magnetoelectric coefficients would be $\alpha_{ij}$.

(ii) The measured magnetoelectric coefficients for a given stress is given by $\tilde{\alpha}_{ij}$ wherein the atoms move when a field is applied.

## ORDERS OF MAGNITUDE

For M and Z-type hexaferrites the crystal structure is hexagonal with space group $P6_3/mmc$. The piezoelectric coefficients $\Lambda^P{}_{ijk}$, the Young's modulus tensor $\varepsilon_{ijkl}$ and the magnetoelastic constants $a_{ijkl}$ have been tabulated [13-17] for similar materials with the same group structure. If the remanence magnetization is along the z-axis (the c-direction) of the hexagonal structure, then the magnetic susceptibility has the approximate diagonal form[16-17]

$$\chi^M = \begin{pmatrix} \chi_\perp & 0 & 0 \\ 0 & \chi_\perp & 0 \\ 0 & 0 & \chi_\| \end{pmatrix},$$

leading to a similar diagonal form for the magneto-electric coefficients

$$\Delta\alpha = \begin{bmatrix} \Delta\alpha_\perp & 0 & 0 \\ 0 & \Delta\alpha_\perp & 0 \\ 0 & 0 & \Delta\alpha_\| \end{bmatrix},$$

where $\Delta\alpha_\perp = 2(\lambda_A - \lambda_B)d_5 C_{44}\chi_{11}/M_R$, and $\Delta\alpha_\parallel = -2(\lambda_B d_1 C_{13} + \lambda_C d_2 C_{33})\chi_{33}/M_R$. The subscripts ($\perp$) and ($\parallel$) imply M perpendicular and parallel to the c-axis. Another simplification or approximation that we have made is that the diagonal $\chi$'s are much higher valued than the off-diagonal elements. Reasonable values of the above parameters may be obtained from literature [13-17] except for the d's and $\lambda$'s for they have not been measured on ME hexaferrites. Y-type hexaferrites are characterized as being space group $R\overline{3}m$ symmetry and the above derivation applies for space group $P6_3/mmc$ only. Nevertheless, we are quoting the following values for the above parameters based on measurements of "similar" oxide materials [13-17]. Thus, we quote the following values $\lambda_A = -15\times10^{-6}; \lambda_B \approx \lambda_C \approx \lambda_D \approx -\lambda_A; d_5 \approx -11\times10^{-12} m/v; d_1 \approx 0.5 d_5 = -0.5 d_2; M_R = 100 a/m$; $C_{11} \approx C_{33} = 1.5\times10^{11} Nt/m^2; C_{13} = 0.6 C_{11} = 1.2 C_{44}; \chi_\perp \approx \chi_\parallel \approx 30$.

Values of $\Delta\alpha$ are listed in table I for single crystal (s), polycrystalline (p), thin films (f) of hexaferrites and composite samples.

Table 1- Calculated and measured α

|  | $\tilde{\alpha}(meas) - s/m$ | $\Delta\alpha(calc) - s/m$ | Reference |
|---|---|---|---|
| Z-type(s) | $2.3\times10^{-6}$ | $0.24\times10^{-6}$ | [5] |
| Z-type(p) | $7.6\times10^{-10}$ | $\sim 10\times10^{-10}$ | [4] |
| M-type(p) | $2.4\times10^{-10}$ | $\sim 10\times10^{-10}$ | [1] |
| M-type(F) | $60.7\times10^{-10}$ | - | [2] |
| Composite | $0.7\times10^{-10}$ | $\sim 10^{-7}$ | [18] |

One may convert above values of $\Delta\alpha$ into CGS units by multiplying above values by $c/4\pi$, where c is the velocity of light. We estimate a $\alpha$ value of $0.24\times10^{-6} s/m$ for the case that the electric field is applied in the z-direction or parallel to the c-axis. As it is well known [13-17] that for polycrystalline samples of Z-type hexaferrite values of $\lambda$'s and d's are considerably reduced, and we estimate roughly $\Delta\alpha \approx 10^{-9} s/m$. We have also simplified the calculation of $\Delta\alpha$ to estimate it for ME composites. We approximated $\Delta\alpha$ for a composite as follows

$$\Delta\alpha = d\lambda C\chi/M .$$

Typically, the piezoelectric strain constant, d, of PZT is $200\times10^{-12}$ m/v [18], $\lambda = 25\times10^{-6}$ [18], C= $1.2\times10^{11} Nt/m^2$ [16-17], $\chi$=10 [17] and M=24,000 a/m [16-18]. The comparison in table 1 is remarkable in view of the fact that we used conservative estimates for all the parameters. Quantitative estimates of the intrinsic ME coupling may not be possible now due to the uncertainty in the parameters. However, this exercise points to a very important result and that is: $\Delta\alpha$ depends on the product of the piezoelectric and magnetostriction constants in the simplest form and this result is new.

## CONCLUSIONS

The thermodynamic arguments presented here show that the linear coupling between the electrical and magnetic systems is best described by a tensor which is proportional to the product of the piezoelectric and magnetostriction tensors. The added feature in hexaferrites is that internal strains induce spin spiral configurations whose cone angle are proportional to this product. The agreement between theory and experiments is reasonable in view of the fact that for hexaferrites the C's and χ' s are known to within 10-20% accuracy, but the d's and λ's can only be estimated within a factor of 10 or more. There are no measurements of these parameters reported in the literature. Our estimate for the composite case is too simplistic, since there was no attempt to account for non-uniformities in the internal strain. For example, the effective magnetostriction is in fact reduced by orders of magnitude due to the soft glue which binds together composite layers. The authors wish to thank the NSF for support.